# Bistable scattering of nano-silicon for super-linear super-resolution imaging

Po-Hsueh Tseng[1,2,†], Kentaro Nishida[2,†,*], Pang-Han Wu[2], Yu-Lung Tang[2], Yu-Chieh Chen[2], Chi-Yin Yang[3], Jhen-Hong Yang[4], Wei-Ruei Chen[4], Olesiya Pashina[5], Mihail Petrov[5], Kuo-Ping Chen[6,*], and Shi-Wei Chu[2,7,8,*]

[1]Graduate Institute of Applied Physics, National Taiwan University, 1, Sec 4, Roosevelt Rd., Taipei 10617, Taiwan

[2]Department of Physics, National Taiwan University, 1, Sec 4, Roosevelt Rd., Taipei 10617, Taiwan

[3]Institute of Imaging and Biomedical Photonics, National Yang Ming Chiao Tung University, 301 Gaofa 3rd Road, Tainan 711, Taiwan

[4]Institute of Photonic System, National Yang Ming Chiao Tung University, 301 Gaofa 3rd Road, Tainan 711, Taiwan

[5]Physics and Engineering Department, ITMO University, St. Petersburg, 197101, Russia

[6]Institute of Photonics Technology, National Tsing Hua University, 101, Section 2, Kuang-Fu Road, Hsinchu, Taiwan 30013, R.O.C.

[7]Molecular Imaging Center, National Taiwan University, 1, Sec 4, Roosevelt Rd., Taipei 10617, Taiwan

[8]Brain Research Center, National Tsing Hua University, 101, Sec 2, Guangfu Road, Hsinchu 30013, Taiwan

*Corresponding author: swchu@phys.ntu.edu.tw; knishida@phys.ntu.edu.tw, kpchen@nctu.edu.tw;

[†] These authors contributed equally to this manuscript.

ABSTRACT

Optical bistability is fundamental for all-optical switches, but typically requires high-Q cavities with micrometer sizes. Through boosting nonlinearity with photo-thermo-optical effects, we achieve bistability in a silicon Mie resonator with a volume size of $10^{-3}$ μm$^3$ and Q-factor < 10, both are record-low. Furthermore, bistable scattering naturally leads to large super-linear emission-excitation power dependence, which we applied to enhance optical resolution by more than 3 times. Our work paves the way toward nanoscale photonics computation and label-free semiconductor nano-inspection.

CONTENT

Semiconductor nanophotonics has received extensive attention in the recent decade, with a particular emphasis on optical nonlinearity, which is critical toward functional photonic circuits. In particular, nano-silicon, as a metamaterial Mie resonator, is emerging rapidly [1,2]. Various nonlinear responses have been unraveled, and the most common form exhibits super-linear power dependency, such as the square or cubic excitation dependence in multiphoton processes [3], leading to applications on all-optical switching [4], super-resolution microscopy [5–7], etc. In principle, the modulation depth of optical switching and spatial resolution of super-resolution microscopy are better with higher orders of nonlinearity. However, for multiphoton excited emission, the order of nonlinearities is mostly two or three, and the emission signals become extremely weak when increasing the order of nonlinearity [8]. Therefore, simultaneous high-order and efficient nonlinear interactions with photons in semiconductor nanostructures are key factors toward practical nanophotonic applications.

To boost optical nonlinear response, with a particular interest in super-linearity, one potential candidate is optical bistability, which contains a vertical state transition in input versus output power dependence [9–11]. In general, optical bistability is produced through the combination of a resonator cavity, and a nonlinear medium whose refractive index depends on incident light intensity, as shown in Fig.1(a). When light enters the cavity, a strong optical field is produced inside the cavity due to resonance. The optical field inside the cavity ($I_{inside}$) modulates the refractive index of the nonlinear medium ($n$), and subsequently the index variation modifies $I_{inside}$. As a result of the iterative feedback between $I_{inside}$ and $n$, the relationship between input ($I_{input}$) and output ($I_{output}$) light intensities of the cavity follows an S-shaped graph [9–11], as shown in the right side of Fig. 1(a). In the "bistable region", $I_{output}$ has multiple values for a single $I_{input}$ value that corresponds with two stable states (yellow dots) and one unstable point (purple dot). To visualize the bistable transition, let's consider that $I_{input}$ gradually increases from zero, initially $I_{output}$ steadily increases in the stable state 1 branch. When $I_{input}$ reaches the critical point, $I_{output}$ increases dramatically to the stable state 2 branch. Now in the upper stable state, when $I_{input}$ decreases, $I_{output}$ remains in the branch until $I_{input}$ reaches another lower critical point to induce a downward transition, creating a hysteresis curve, which is a hallmark of bistability.

For silicon-based optical bistable systems, they require high-Q cavities, such as photonic crystal or ring cavity, whose Q-factors range from $10^3$ to $10^6$ [12,13], as shown in Fig. 1(b), to compensate the low Kerr nonlinearity in silicon ($n_2 \sim 10^{-9}$ μm$^2$/mW) [1]. However, all the bistable resonators have cavity sizes at least in the micrometer scale. Although some unique optical effects in the coupled nonlinear subwavelength nanoresonators and discrete waveguides such as the appearance of solitonic and oscillonic regimes and optical bistability [14–17], but they have not been experimentally demonstrated due to the requirement of strong optical field. For nanoscale silicon, the Mie resonator volume could be a few orders smaller than the above resonators, but the Q-factor of common silicon Mie resonance is only around 10. To enable bistability in nano-silicon, possible approaches are enhancing either resonance Q-factor through structural design [18], or the nonlinearity magnitude. Recently, we discovered that photothermal and thermo-optical effects in a silicon Mie resonator could induce giant nonlinear optical responses, with an equivalent $n_2$ value reaching 1 μm²/mW, several orders larger than that of bulk [19–22]. In principle, this large nonlinearity shall relax the requirement of the high-Q cavity.

In this work, we demonstrate optical bistability in a silicon Mie resonator with Q-factor ~ 10, and $10^{-3}$ μm³ volume size, which are both significantly smaller compared with other representative silicon-based bistable devices made with photonic crystals [23–25] or ring cavities [12,13,26–28], as shown in the comparison chart of Fig.1(c) (See Supplementary Materials Table S1 for more detailed comparison). We established a photothermal model that explains the emergence of optical bistability, whose underlying mechanism is temperature-dependent nonlinear absorption due to the interplay between Mie resonance and thermo-optical effect. Under a laser scanning microscope (LSM), hysteresis of optical scattering from a single silicon nanostructure is identified, leading to excitation-scattering intensity dependency as large as 10th power. Via the exceptionally large super-linearity, we feature dramatic reduction of point spread function (PSF) width and improvement of spatial resolution in LSM.

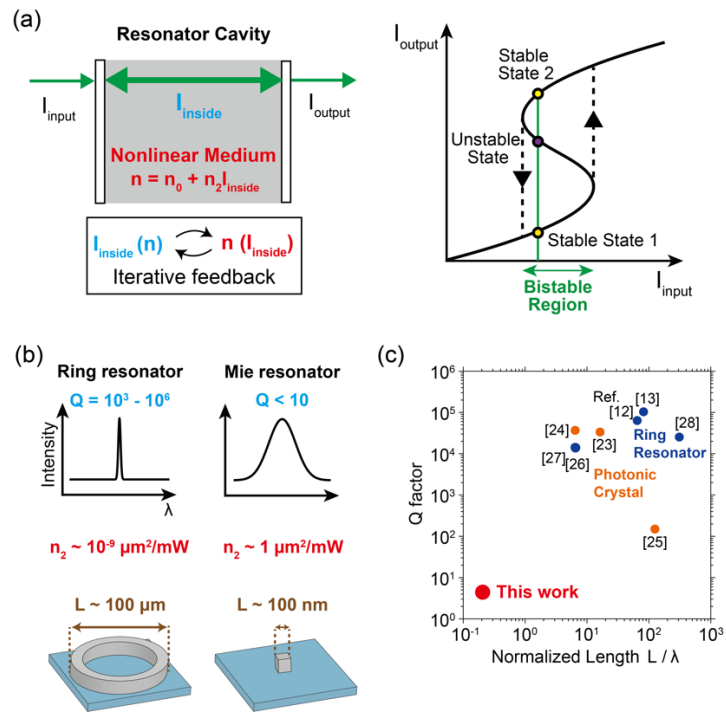

Fig. 1. (a) Principle of optical bistability. The left schematic of a resonator cavity where $I_{input}$ is input intensity, $I_{inside}$ is the field intensity inside the cavity, $I_{output}$ is output intensity. The cavity is filled with a nonlinear medium with the refractive index ($n$) dependent on $I_{inside}$ as $n=n_0+n_2 I_{inside}$, where $n_0$ and $n_2$ are linear and nonlinear refractive index coefficients. The right schematic shows a S-shaped $I_{output}$ - $I_{input}$ dependency, where a bistable region is marked. The yellow dots are stable states, while the purple dot represents the unstable state. (b) Comparison between silicon ring resonator and our Mie resonator, for Q-factor (Q), the effective nonlinear coefficient ($n_2$) and footprint size of the structure (L). (c) Comparison of resonance Q-factor and normalized footprint over resonant wavelength (L/λ) for various silicon-based bistable resonators: photonic crystals (orange), ring resonators (blue) and our Mie resonator.

In order to achieve steep super-linear scattering, we exploited photothermal optical bistable states of silicon nanoparticles. Fig. 2(a) illustrates the concept of photothermal bistability, which is governed by competition between photothermal heating (green arrows) and heat dissipation (purple arrows). The green line indicates the heating strength, which is dominated by the temperature-dependent absorption cross section ($\sigma_{in}$) of silicon nanoparticles. The reason that the green line exhibits a peak is because of the Mie resonance redshift at elevated temperature. The excitation wavelength shall be on the red side of a resonance peak. As the temperature grows, the absorption peak shifts to red-side due to thermo-optical effect, so $\sigma_{in}$ sequentially increases and decreases. Without losing generality, here we use a Gaussian peak to approximate $\sigma_{in}$ thermal dependence. On the other hand, the purple line indicates the heat diffusion cross section ($\sigma_{out}$) from silicon nanoparticles. Based on the Poisson equation, the heat diffusion proportionally increases versus the temperature. The three intersections of the green and purple lines represent thermal equilibrium solutions ($\sigma_{in} = \sigma_{out}$), but their stabilities rely on temperature convergence. The black intersection dot in Fig. 2(a) is an unstable state, since it is located at the transition from a blue region ($\sigma_{in} < \sigma_{out}$, cooling) to an orange region ($\sigma_{in} > \sigma_{out}$, heating), resulting in temperature divergence under perturbation. On the contrary, the two yellow dots are stable points, because if the local temperature fluctuates low to the orange region ($\sigma_{in} > \sigma_{out}$, heating), or high to the blue region ($\sigma_{in} < \sigma_{out}$, cooling), the thermal flux brings it back to the stable state. Therefore, a silicon Mie resonator has two thermally stable states under one excitation condition, i.e. the hallmark of bistability, and the corresponding optical response shall change dramatically during state transitions.

The physical picture of bistability is more explicitly given in Fig. 2(b), via defining a generalized potential $U(T)$ [11] as differentiating the total heat flux (i.e., $\sigma_{in}$ - $\sigma_{out}$) with respect to the temperature $T$,

$$-\frac{\partial U}{\partial T} = \sigma_{in} - \sigma_{out} \qquad (1)$$

Apparently, the generalized potential has two local minima and one local maximum, which respectively correspond to the two stable states and one unstable state.

We theoretically and experimentally confirmed the existence of bistability for an amorphous silicon nanocuboid, with 155 nm width and 142 nm height as shown in the scanning electron microscopy (SEM) image (Fig. 2(c)). The fabrication details follow our previous research [30]. Fig. 2(d) shows the thermal equilibrium simulation of amorphous silicon nanocuboids with ~150 nm width and height (see Supplementary Materials for more details on simulation conditions). The solid line is temperature-dependent absorption cross section of amorphous silicon nanocuboids by solving electromagnetic Mie equations through finite element method (FEM) [31], and the dotted lines are solutions of the Poisson heat equation under different excitation intensities, whose slope reduces with increasing excitation intensity. The thermal equilibrium states exist at the intersections of dotted and solid lines. When the excitation intensity is lower than 1.98 mW/μm², only one intersection is found, such as the point A in the figure, and thus the silicon nanocuboid has a single thermal equilibrium state. When the intensity reaches 1.98 mW/μm², two intersection points emerge (point B and F). Intriguingly, in the excitation intensity range between 1.98 and 3.05 mW/μm², there are three intersections, where the central one is unstable, as explained in Fig. 2(a). Therefore, here the silicon nanocuboid has two thermally stable states, i.e. bistability. Now as the excitation intensity reaches 3.05 mW/μm² and above, the intersection points reduce to two (C and D) and one (E) again, respectively, manifesting that the number of equilibrium states comes back to unity at the elevated incident condition.

Based on the results of Fig. 2(d), we plot photothermal bistability in Fig. 2(e) through explicit excitation versus scattering intensity dependence. When the excitation intensity is lower than 1.98 mW/μm$^2$, the silicon nanocuboid has a single thermal equilibrium state, and thus scattering intensity moves along the A-B line. When the excitation intensity is between 1.98 and 3.05 mW/μm$^2$, the silicon nanocuboid goes into the bistable region. However, without state transition, the silicon nanocuboid stays in the lower branch, i.e. the B-C segment, of this bistability region. If the excitation intensity exceeds 3.05 mW/μm$^2$, silicon nanocuboid reaches the end of the bistable region (point C) and the thermal state jumps to the upper branch (D-E segment), accompanied with an abrupt increase of scattering intensity. The E-F segment marks the condition that the silicon nanocuboid stays at the upper branch when the excitation intensity decreases from 3.05 mW/μm$^2$. The downward thermal state transition occurs at point F to point B, thus forming the photothermal hysteresis loop.

In Fig. 2(e), the downward transition of the F-B segment is smaller than that of the upward transition of C-D segment, even though in Fig. 2(d), the temperature variations of these two segments are both around 500-600 K. This is because the scattering cross section variation over temperature is not linear. Our calculation predicts the existence of hysteretic scattering intensity, with huge super-linearity near the upward/downward transition, from a single amorphous silicon nanocuboid.

We experimentally measured the scattering response from nanocuboid by dark-field LSM, equipped with a 785 nm laser [19] (see Supplementary Materials and Fig. S1). The laser wavelength is located on the red side of a Mie absorption peak, a prerequisite to form bistability (Supplementary Materials Fig. S2). The numerical apertures of focusing objective and collection condenser are 0.95 and 1.4, respectively. Fig. 2(f) is the log-log power dependency of scattering while excitation intensity is modulated through the combination of a λ/2 waveplate and a polarization beam splitter. When the excitation intensity increases (blue curve), the steep transition appears at ~3 mW/μm$^2$, agreeing well with the simulation of Fig. 2(e). The log-log slope of the transition, i.e. the $p$ value that represents the order of super-linearity, is around 10. Note that before and after the transition, the slope dramatically changes, manifesting the existence of two distinct thermally stable states. On the other hand, when the excitation intensity decreases from the upper branch (red curve), the steep transition occurs at a lower excitation intensity, featuring clear hysteresis response and demonstrating the existence of bistability.

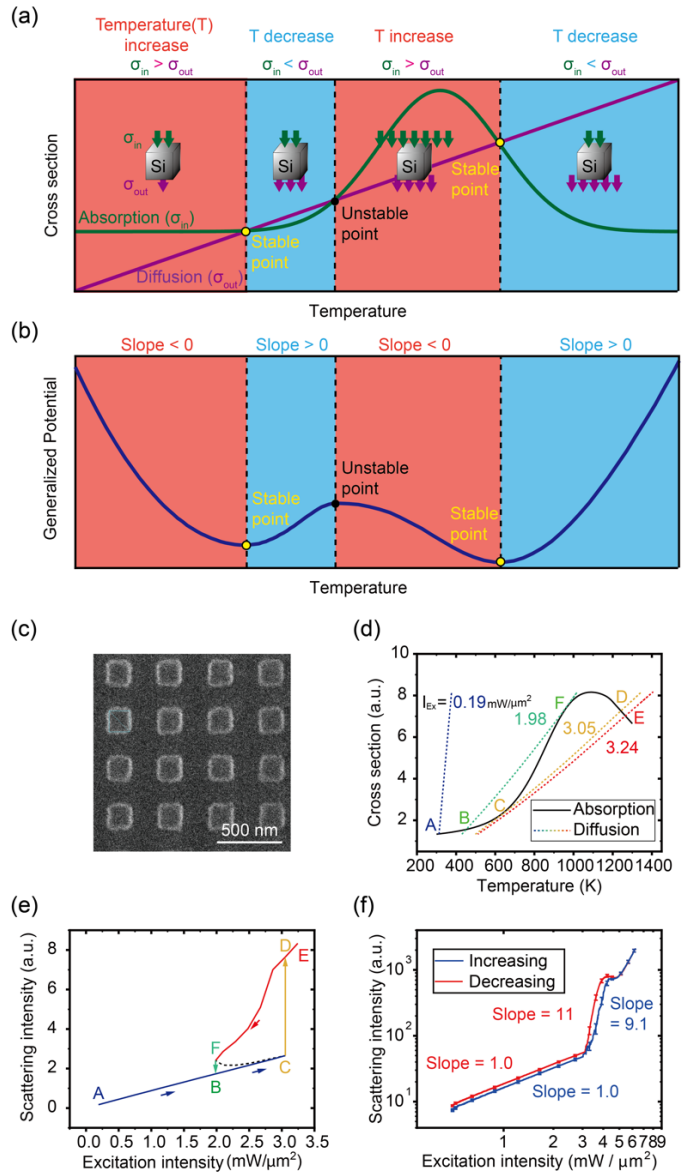

Fig. 2. (a) Conceptual illustration of photothermal optical bistability in a silicon Mie resonator. Green and purple lines respectively indicate heat influx ($\sigma_{in}$) via absorption, which is nonlinear due to thermo-optical effect, and heat outflux ($\sigma_{in}$) from diffusion, calculated from Poisson equation. Orange and blue regions represent the conditions of $\sigma_{in} > \sigma_{out}$ and $\sigma_{in} < \sigma_{out}$, respectively. The intersection points are thermal equilibrium solutions, and the two yellow dots are stable states, while the black dot indicates an unstable state. (b) Bistability visualized through the generalized potential diagram. (c) SEM image of amorphous silicon nanocuboids sample. (d) The solid curve is temperature-dependent absorption cross section calculated by the FEM method. The dotted lines are heat diffusion cross sections derived from the Poisson heat equation at various excitation intensities. Points A to F are intersections between the solid line and the dotted lines, i.e. thermal equilibrium solutions. (e) Simulated hysteresis scattering, where points A to F correspond to those intersections in (d). (f) Hysteresis by fixed focal spot measurement of scattering intensities from a single silicon nanocuboid. Blue and red curves are obtained by increasing and decreasing the excitation intensities, respectively, at a fixed focal point. The error bar indicates the standard error.

It is known that super-linear scattering directly improves the spatial resolution of LSM without additional hardware or software post processing [5], and the reason is briefly explained here. As shown in Fig. 3(a), in LSM, an image is acquired by scanning the sample with a focused laser beam, whose intensity distribution $I_{exc}(x)$ follows a Gaussian function,

$$I_{exc}(x) = A exp(\frac{x^2}{2x_0^2}) \tag{2}$$

where $x$ is the lateral position, $A$ is the peak intensity, and $x_0$ is the width. When the emission intensity $I_{emi}$ has $p^{th}$ order super-linear power dependence over excitation intensity $I_{exc}$, i.e. slope "$p$" in a log-log plot, the emission PSF is expressed as:

$$I_{emi}(x) \propto (I_{exc}(x))^p = (A exp(\frac{x^2}{2(x_0)^2}))^p = A' exp(\frac{x^2}{2(\frac{x_0}{\sqrt{p}})^2}) \tag{3}$$

That is, the size of PSF proportionally reduces to the square root of $p$, leading to remarkable resolution enhancement when $p$ is large. The idea is schematically presented in Fig. 3(b). The blue dot and the blue box show that when $I_{emi}$ linearly responds to $I_{exc}$, i.e. $p = 1$, the emission PSF follows $I_{exc}(x)$. When super-linearity exists, i.e. $p > 1$ at the green dot, the green box illustrates effective resolution enhancement. Our interest lies in boosting the super-linearity, i.e. $p \gg 1$, and thus significantly enhancing optical resolution, as shown by the red dot and red box.

To verify the application of super-linear scattering to super-resolution imaging, Fig. 3(c) shows a schematic LSM imaging process and the resulting image cross-sectional profiles. When the focal spot approaches the nanocuboid, as shown in the left panel, excitation intensity escalates to the upward transition intensity $I_{up}$ and induce the scattering state transition from the hysteretic lower branch to upper branch, as indicated by the inset. The dotted line presents the corresponding image profile, where the upward super-linear transition is expected. In the middle panel of Fig. 3(c), when the focal spot just passes through the center of a nanocuboid, excitation intensity starts to decrease and scattering intensity arrives at the green segment of the middle inset, i.e. the hysteretic upper branch. Due to hysteresis, the emission stays at the upper branch while excitation reduces from $I_{up}$ to $I_{down}$, corresponding to the green dots in the image profile. In the right panel, the focal spot leaves the nanocuboid, and excitation intensity decreases over the downward transition intensity $I_{down}$. During the downward state transition, as shown by the red line in the inset and dots in the profile, the scattering decreases fast, leading to an overall reduction of full-width-at-half-maximum, i.e. resolution enhancement. Notably, the LSM image profile should be asymmetric because the thresholds of state transition are different when the focal spot approaching and leaving the nanocuboid.

The corresponding results are presented in Fig. 3(d) and 3(e), where raster scanning is enabled through a set of galvanometer mirrors. Due to diffraction limit, the spatial resolution without bistability is ~500 nm, and thus we use a silicon nanocuboid array of ~400 nm periodicity to experimentally demonstrate the improvement of LSM spatial resolution. At a low excitation intensity of 2.2 mW/μm$^2$, where the scattering is linear, as shown in Fig. 3(d) and the inset, the structure of the silicon nanocuboid array is not visible. At an elevated excitation intensity, Fig. 3(e) demonstrates that the array structure is clearly resolved, with full-width-at-half-maximum around 140-nm, manifesting more than 3-fold resolution enhancement via bistability. Interestingly, the signal profile indeed reveals asymmetry due to the hysteresis effect, as we expected. Note that in Fig. 3(e) the images are acquired by laser scanning, while in Fig. 3(b) the hysteresis power dependency is taken with a fixed focal spot, resulting in slightly different bistability threshold probably due to less heat accumulation with the moving spot.

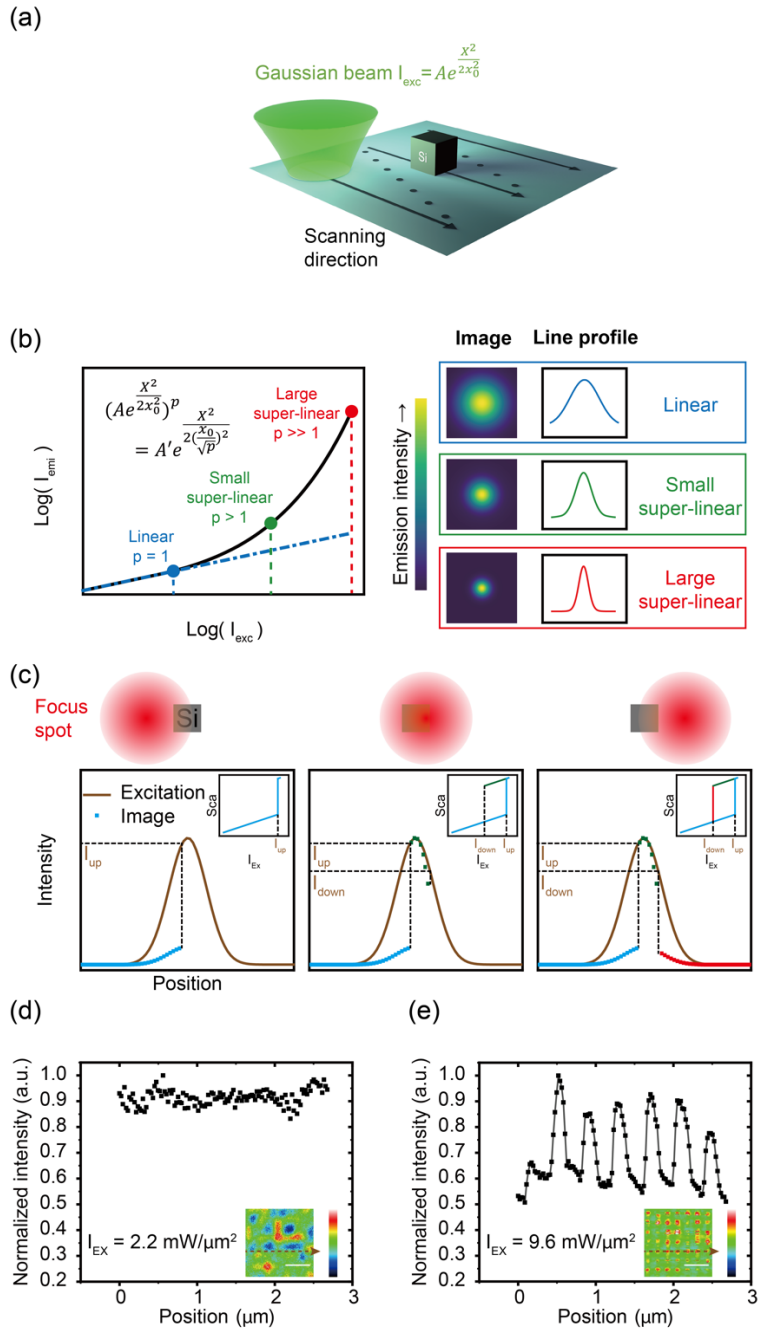

Fig. 3. (a) Schematic of lateral scanning process in LSM. (b) The connection between excitation/scattering intensity dependence and resolution enhancement, showing linear (blue), small super-linear (green) and large super-linear (red) cases. (c) Schematic of laser scanning imaging with super-linear hysteresis scattering. The red circle and the gray cube indicate the focal spot of the excitation beam and a silicon nanoparticle, respectively. The three figures show line profiles of excitation (brown curves) and scattering intensities (dots). Insets indicate the scattering response in a hysteresis scattering curve. (d,e) LSM signal profiles of the silicon nanoarray measured at the excitation intensities of 2.2 mW/μm$^2$ (d) and 9.6 mW/μm$^2$ (e), and the corresponding signal profiles obtained at the dotted arrow of the scattering images. The scattering images are shown in the insects. The scale bars in the image indicate 1 μm.

In this paper, we demonstrated that photo-thermo-optical bistability exists with a Mie resonator of diameter ~100-nm and Q-factor ~ 10. The transition between the bistable states offers particularly steep super-linearity of Rayleigh scattering, experimentally reaching 10th power of excitation intensity. By applying the super-linear scattering to LSM, we successfully resolved closely aligned subwavelength nanoarray structures. Compared with conventional super-resolution imaging technique for silicon nanostructures imaging, such as pump-probe [32], subtraction [33], saturated excitation (SAX) [19], donut suppression [34], algorithm calculation [35], etc., our approach has the advantage of the simplicity of the imaging system: no requirements of add-on hardware or post-processing. Our research not only helps to understand the physical mechanism of nonlinear heating/scattering in silicon nanostructures, but also points out the direction toward designing the optimal nanostructure for largest nonlinearity, especially for applications such as all-optical switching, and nano-probes of super-resolution microscopy.

FUNDING

Ministry of Science and Technology Taiwan, MOST111-2119-M-002-013-MBK, MOST111-2321-B-002-016, MOST109-2112-M-002-026-MY3, NTU-111L8809, The Featured Areas Research Center Program within the framework of the Higher Education Sprout Project co-funded by the MOST and the Ministry of Education, Taiwan (MOE).